\documentclass[]{iopart}

\usepackage{iopams}
\usepackage{graphicx}
\usepackage{cite}
\usepackage{color}

\definecolor{red}{rgb}{1,0,0}

\newcommand{\bm}[1]{\mbox{\bf {\em #1}}}

\graphicspath{{./}{./plots_renamed/}}

\begin{document}

\title[Semiflexible polymer in microcapillary flows]
{Semiflexible polymer conformation, distribution and migration
in microcapillary flows}

\author{Raghunath Chelakkot$^{\dag}$, Roland G. Winkler$^\ddag$, Gerhard Gompper$^{\dag, \ddag}$}

\address{$^\dag$Institut f\"ur Festk\"orperforschung and $^{\ddag}$Institute for Advanced
Simulation,  Forschungszentrum J\"ulich,
  52425 J\"ulich, Germany\\}
\begin{abstract}
The flow behavior of a semiflexible polymer in microchannels is
studied using Multiparticle Collision Dynamics (MPC), a
particle-based hydrodynamic simulation technique. Conformations,
distributions, and radial cross-streamline migration are
investigated for various bending rigidities, with persistence
lengths $L_p$ in the range $0.5 \le L_p/L_r \le 30$. The flow
behavior is governed by the competition between a hydrodynamic
lift force and steric wall-repulsion, which lead to
migration away from the wall, and a locally varying flow-induced
orientation, which drives polymer away from the channel center and
towards the wall. The different dependencies of these effects on
the polymer bending rigidity and the flow velocity results in a
complex dynamical behavior. However, a generic effect is the
appearance of a maximum in the monomer and the center-of-mass
distributions, which occurs in the channel center for small flow
velocities, but moves off-center at higher velocities.

\end{abstract}

\maketitle

\section{Introduction}

\subsection{Polymers, Vesicles and Cells in Micro- and Nanochannels}

Complex fluids --- such as polymer solutions, colloidal
dispersions, and suspensions of vesicles or cells --- exhibit an
intriguing flow behavior, particularly in confined geometries. The
interplay of internal degrees of freedom, e.g, conformational
changes of polymers or shape changes of vesicles and cells,
combined with fluid mediated interactions, inhomogeneous flow
profiles, and wall interactions lead to novel and often unexpected
effects
\cite{wall:78,broc:77,vlie:90,jend:03,bald:07,jend:03_1,tege:04,agar:94,jend:04,gg:gomp05g,usta:06,khar:06,stei:06,usta:07,cann:08,mark:09,gg:gomp09d,gg:gomp10d}.
Insight into theses aspects provides not only an understanding of
phenomena for conventional applications such as lubrication,
adhesion, polymer processing, and blood flow, but also of emergent
microfluidic devices \cite{squi05,whit:06} and the lab-on-a-chip
technologies \cite{crai06} with their micro- or nano-size length
scale. In this context, the investigation and analysis of single
molecules or cells is of particular interest, since on the one
hand such knowledge is essential to develop and optimize micron-
and nano-scale devices, while on the other hand the study of
individual particles often provides a more detailed microscopic
knowledge than can be extracted from an ensemble.

The size of red blood cells (RBCs), about $8\mu$m in diameter, is
comparable to the size of microvessels and microcapillaries in
a mammalian body. Thus, the deformability and dynamics of an
individual cell determines the flow behavior of blood in the
capillary network. Due to the physiological importance of this
process and the relatively large size of the objects, detailed
investigations of this system started already about 50 years ago
\cite{skal69,seco86,prie96,pozr05a}. Modern microfluidic
techniques \cite{blaz:06,whit:06,utad:07} and new mesoscale
simulation approaches \cite{dzwi03,gg:gomp04h,dupi07,gg:gomp08d}
provide new insight into the delicate interplay between flow
forces, cell deformation, and hydrodynamic interactions
\cite{gg:gomp05g,gg:gomp09d,abka08b,toma09}. In particular, in
microchannels slightly larger than the RBC diameter, it has been
found that single red blood cells transform from their equilibrium
shape of biconcave disks into a parachute shape at a critical flow
velocity, which depends linearly on the bending rigidity and the
shear modulus of the cell membrane \cite{gg:gomp05g}. At small
volume fractions of red blood cells, they show a strong clustering
tendency due to hydrodynamic interactions \cite{gg:gomp09d}. At
higher volume fractions, several distinct phases are found, which
range from disordered arrangement of discocytes over single-file
motion of aligned parachutes to a staggered, zig-zag arrangement
of slipper shapes \cite{gg:gomp09d}. In channels which are smaller
than the RBC diameter, they assume bullet shapes at
sufficiently high flow velocities \cite{seco86,abka08b,toma09}.

Fluid vesicles are distinct from red blood cells since their
membrane is fluid, and therefore has a vanishing shear modulus.
The flow behavior of vesicles has been studied intensively in
recent years, in particular in shear flows. Much less is known
about fluid vesicles in capillary flows. In narrow, homogeneous
channels, vesicles assume bullet shapes \cite{vitk04}. In
structured, saw-tooth shaped channels, which are somewhat larger
than the vesicle size, a complex conformational and dynamical
behavior has been found \cite{gg:gomp10d}. For nearly spherical
vesicles, a transition from symmetric shape oscillations to
orientational oscillations has been predicted with decreasing flow
velocity. For shapes, which deviate more strongly from a sphere,
experiments and simulations show new shapes with two symmetric or
a single asymmetric tail \cite{gg:gomp10d}.

In narrower channels, with a diameter on the order of 100 nm,
it is possible to investigate the static and dynamic behavior of
single DNA molecules. This system is interesting for several
reasons. First, microfluidic devices allow for the manipulation,
sizing, and sorting of DNA fragments \cite{chou99,levy10}, and
thereby a direct visualization of genomic information
\cite{tege:04,pers10}. Second, DNA has turned out to be an ideal
model to study the properties of single polymers, because, when
fluorescently labelled, its conformations, dynamics, and flow
properties can be observed directly under a microscope
\cite{wall:78,perk95,smit99,reis:05,schr:05,gera:06} or by
correlation spectroscopy \cite{petr:06}. Thus, DNA is also very
well suited to study the behavior of semiflexible polymers in
micro- and nanochannels.

With a persistence length of $L_p \simeq 17 \mu m$,
filamentous actin is a biopolymer for which semiflexible
behavior becomes important already in wider channels than for DNA
(with its persistence length $L_p \simeq 50 nm$). Thus,
experiments with actin are essential to elucidate the behavior of
semiflexible polymers in microchannels without
\cite{koes05,koes08,koes09} and with flow \cite{stei:08}. In
equilibrium, the parallel extension of a semiflexible polymer
shows scaling behavior with power-law dependencies on channel
diameter and persistence length
\cite{odij83,reis:05,gg:gomp07e,levi07,odij08}.

Structured channels are particular interesting, because they allow
to study time-dependent flows \cite{gg:gomp10d}, threshold forces
or flow rates for penetration into narrow pores
\cite{gg:gomp95h,saka05,link06,mark:09}, and flow injection into a
pore \cite{mark:09}. For fluid vesicles, it has been shown that
the threshold strength of a driving field (e.g., an electric
field) for narrow pores increases nearly linearly with membrane
bending rigidity and vesicle area, and decreases rapidly with the
pore radius \cite{gg:gomp95h}. For polymers, it has been predicted
by scaling arguments \cite{saka05} and confirmed by simulations
\cite{mark:09} that the threshold velocity flux for entry into a
narrow pore is {\em independent} of both the polymer length and
the pore radius.

\subsection{Polymer Migration in Micro- and Nanochannels}

An interesting aspects of polymer transport in the presence of a
zero-slip wall is cross-streamline migration. This effect has been
observation in planar shear flow in the presence of walls as well
as Poiseuille flows
\cite{jend:03,jend:03_1,jend:04,usta:06,khar:06,usta:07,cann:08,send:08}
and in experiments~\cite{tege:04,stei:06,bald:07}, as a formation
of depletion layer near the wall. The thickness of the depletion
layer is found to increase with flow, which indicates that
migration away from the wall increases with the flow rate. This
migration phenomenon is explained as a results of polymer-wall
hydrodynamic interactions. Analytical as well as simulation
studies, which properly account for such hydrodynamic
interactions, qualitatively reproduced the experimentally observed
wall-induced migration.

In microchannels, confinement effects and a spatially varying
shear rate due to the parabolic flow velocity profile imply
additional features compared to simple shear flows. The steric
interaction with the channel wall restricts the conformations of a
flexible polymer and induce alignment to a semiflexible polymer.
Moreover, the spatially varying local shear rate changes the
conformations and alignment of a polymer as a function of its
lateral position. In addition to the wall hydrodynamic
interactions, these two effects also influence the qualitative
behavior of cross-streamline migration. Simulations of flexible
polymers have shown that under strong confinement there is a net
migration away from the channel center, contrary to the
predictions of wall induced inward migration~\cite{usta:06}. The
outward migration is also found in relatively weak
flow~\cite{cann:08}. In both cases, the outward migration has been
attributed to the suppression of steric interactions with the wall
due to an enhanced alignment by the flow. At large flow strengths,
wall hydrodynamic interactions dominate, resulting a net inward
migration of the polymer center-of-mass. Interestingly, the
center-of-mass density at the channel center is also found to
decrease with flow and at large flow strengths a  density  maximum
is found at a distance away from the center. The reason behind the
off-center peak of the center-of-mass distribution for flexible
polymers is the flow-induced conformational change, which leads to
stretched polymers close to the wall and coiled chains in the
central part
\cite{jend:03,jend:03_1,jend:04,usta:06,khar:06,usta:07,cann:08}.
This give rise to an enhanced outward diffusion and ultimately
leads to a concentration dip at the center.

Although the cross-streamline migration behavior of flexible
polymers has been extensively studied, there are only a few
studies, which focus on the migration behavior of  semiflexible
and rodlike polymers \cite{schi:96,sain:06}. Understanding the
migration properties of semiflexible and stiff polymers is
important for the study of many biopolymers, such as actin
filaments, microtubules, and intermediate filaments. Since
polymer conformations are determined by its rigidity, the
migration behavior of semiflexible and rodlike polymers can differ
significantly from that of flexible polymers.

In this paper, we use computer simulations to study and discuss
the migration properties of semiflexible polymers with persistence
length ranging from half up to many times of its contour length.
We employ a hybrid simulation scheme including Molecular Dynamics
(MD) for the polymers and Multiparticle Collision Dynamics (MPC),
a particle-based hydrodynamics simulation technique, for the
solvent, so that wall hydrodynamic interactions and confinement
effects are taken into account explicitly. The details of the
method are given in Sec.~\ref{sec:methods}. Our results, presented
in Secs.~\ref{sec:concentration} and \ref{sec:alignment}, show
that polymer migration is indeed qualitatively influenced by
bending rigidity. Furthermore, in Sec.~\ref{sec:whi}, we
investigate the importance of hydrodynamic interactions by
comparing the results of simulations with MPC and Brownian
solvents. Finally, the dynamics of polymer migration across the
channel is studied in Sec.~\ref{sec:migration}

\section{Model and Simulation Method}
\label{sec:methods}

\subsection{Semiflexible Polymer}

The linear polymer is comprised of $N_m$ point-like monomers of
mass $M$ each, which are connected by the harmonic potential
\begin{equation}
U_s = \frac{\kappa_s}{2} \sum_{i=1}^{N_m-1} \left(|{\bm r}_{i+1}
-\bm{r}_i | - b \right)^2,
\end{equation}
where  $\bm{r}_i$ denotes the position of particle $i$,  $b$
the bond length, and $\kappa_s$ the force constant.
Excluded-volume interactions are taken into account by the
Lennard-Jones potential \cite{muss:05}
\begin{equation}
U_{LJ} = \left\{ \begin{array}{ll}
  4 \epsilon \left[ \left( \frac{ \displaystyle \sigma}{\displaystyle  r}\right)^{12} -
\left( \frac{\displaystyle  \sigma}{\displaystyle  r}\right)^{6}\right] + \epsilon \ , & r < 2^{1/6} \sigma \\
0 \ , & \mbox{else}
\end{array}  \right. .
\end{equation}
To account for polymer stiffness, the bending potential is
applied~\cite{wink:04}
\begin{equation}
U_b = \frac{\kappa_b}{2} \sum_{i=2}^{N_m-1} \left({\bm R}_{i+1}
-\bm{R}_{i}  \right)^2 ,
\end{equation}
where $\bm{R}_i =\bm{r}_i-\bm{r}_{i-1}$ is the bond vector and
$\kappa_b$ is the bending rigidity. In the semiflexible limit, the
bending rigidity is related to the persistence length by
$L_p=\kappa_b/k_BT$.

The dynamics of the polymer is described by Newtons' equations of
motion, which are integrated using the velocity-Verlet algorithm.

\begin{figure}
\begin{center}
\includegraphics*[width=12cm,clip]{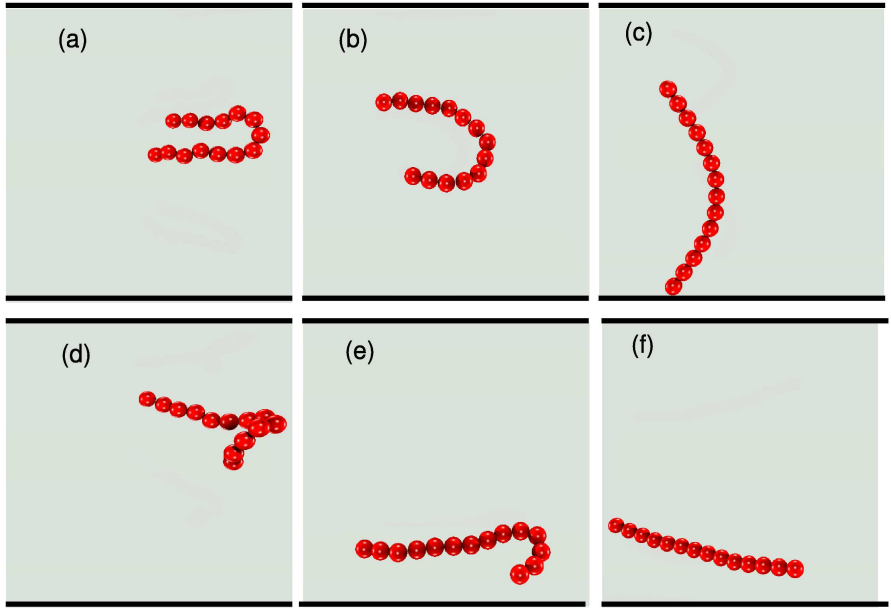}
\caption{Snapshots of polymer conformations close to the channel center (top)
and the wall (bottom) for the Peclet number $Pe =360$ and the 
persistence lengths
$L_p/L_r \approx 0.5$  (a), (d),  $2.1$ (b), (e), and  $30.8$ (c), (f).
Movies are available as Supporting Material \cite{movie}.
\label{fig:snapshot}}
\end{center}
\end{figure}

\subsection{Hydrodynamic Solvent}

In the MPC algorithm, the fluid is described by a set of $N$
point-like particles of mass $m$ each, which move in continuous
space with velocities determined by a stochastic process. Their
dynamics evolves in two steps.  In the streaming step, the solvent
particles move ballistically for a time $h$, which we denote as
collision time. In the collision step, particles are sorted into
the cells of a cubic lattice of lattice constant $a$ and their
relative velocities, with respect to the center-of-mass velocity
of each cell, are rotated around a random axis by an angle
$\alpha$. The orientation of the axis is chosen independently for
every cell and collision step. For every cell, mass, momentum, and
energy are conserved in this process. The algorithm is described
in detail in
refs.~\cite{male:99,male:00,kapr:08,gomp:09,ripo:04,ripo:05}. The
fluid is confined in a cylindrical channel with periodic boundary
conditions along the channel axis. No-slip boundary conditions are
imposed on the channel walls by the bounce-back rule and virtual
wall particles, as described in ref.~\cite{lamu:01}. The flow is
induced  by a gravitational force ($mg$) acting on every fluid
particle.

The interaction of a polymer with the solvent is realized by
inclusion of its monomers in the MPC collision
step~\cite{male:00_1}. Between two MPC steps, several MD steps are
performed to update the positions and velocities of the monomers.
Extensive studies of polymer dynamics confirm the validity of this
procedure~\cite{ripo:05,webs:05,muss:05,male:00_1}.

To maintain a constant temperature, the velocity scaling algorithm
is applied as described in Ref. \cite{huan:10}. Here, an kinetic
energy $E_k$ is chosen from the gamma distribution independently
for every cell, and the individual particle velocities of that
cell are multiplied by the factor $(2 E_k/(m \sum
\bm{v}_i))^{1/2}$, where the sum runs over all particles in the
considered cell. This assures that the velocity distribution of
the particles is Maxwellian. Similarly, a Maxwellian
velocity distribution is obtained by the Monte Carlo procedure
described in Ref. \cite{hech:05}.

Simulations of the pure solvent system yield velocity profiles
which agree with the solution of Stokes' equation for the
considered geometry.

\subsection{Brownian Solvent}

An advantage of the MPC approach is that hydrodynamic interactions
can easily be switched off, without altering the monomer diffusion
significantly \cite{kiku:03,ripo:07}. In this case, denoted as
Brownian MPC, each monomer independently performs a stochastic
collision with a phantom particle with a momentum taken from the
Maxwell-Boltzmann distribution with variance $m \left\langle N_c
\right\rangle k_BT$, where $\left\langle N_c \right\rangle$ is the
average number of solvent particles per collision cell
\cite{ripo:07}.

\subsection{Parameters}

For the solvent, we employ the parameters $\alpha =130 ^\circ$,
$h=0.1 \tau$, with $\tau=\sqrt{ma^2/k_{B}T}$ ($k_B$ is Boltzmann's
constant and $T$ is temperature), $\left\langle N_c \right\rangle
=10$, $M=m \left\langle N_c \right\rangle, b=\sigma =a$, the fluid
mass density $\varrho = \left\langle N_c \right\rangle m/a^3$, and
$k_BT/\epsilon =1$. The time step in the MD simulation is set to
$h_{MD} = 5\times10^{-3} \tau$. A polymer with $N =14$ monomers is
placed in a cylindrical channel of radius $R = 8 a$. With the
length $L_r=(N-1)a=13a$, the polymer does not interact with the
wall when its center of mass is near the channel center. In order
to maintain a constant contour length of the polymer, we set
$\kappa_s = 3 \times 10^3 /(k_BT/b^2)$. The persistence length
$L_p$ is varied, by selecting the values $\kappa_b /(k_BT/b^2) =
7, \ 28, \ 50, \ 100, \ 200, \ 400$, which correspond to the
persistence lengths $L_p/L_r \approx 0.5, \ 2.1, \ 3.8, \ 7.7, \
15.4, \ 30.8$.  The channel length is $28 a$.

The strength of the applied pressure field is characterized by the
Peclet number $Pe = {\dot \gamma} \tau_R$, where ${\dot \gamma} =
g \varrho R /(2 \eta)$ is the shear rate at the cylinder wall. The
Reynolds number $Re = \varrho R v_m/ \eta = \varrho R^2 \dot
\gamma/(2 \eta)$, where $v_m$ is the maximum fluid velocity,
depends linearly on the shear rate. For the above MPC parameters,
the viscosity \cite{ripo:04} is such that $Re < 1$ for all
considered $\dot \gamma$. Equilibrium simulations for a system
with periodic boundary conditions yield the end-to-end vector
relaxation time $\tau_R \approx 3200 \tau$. This value agrees
within approximately $20\%$ with the relaxation time obtained
theoretically for a semiflexible polymer with the same ratio
$L_r/L_p$ \cite{wink:06_1}. The relaxation time of the Brownian
MPC simulation is $\tau_R \approx 8200 \tau$.

\begin{figure}
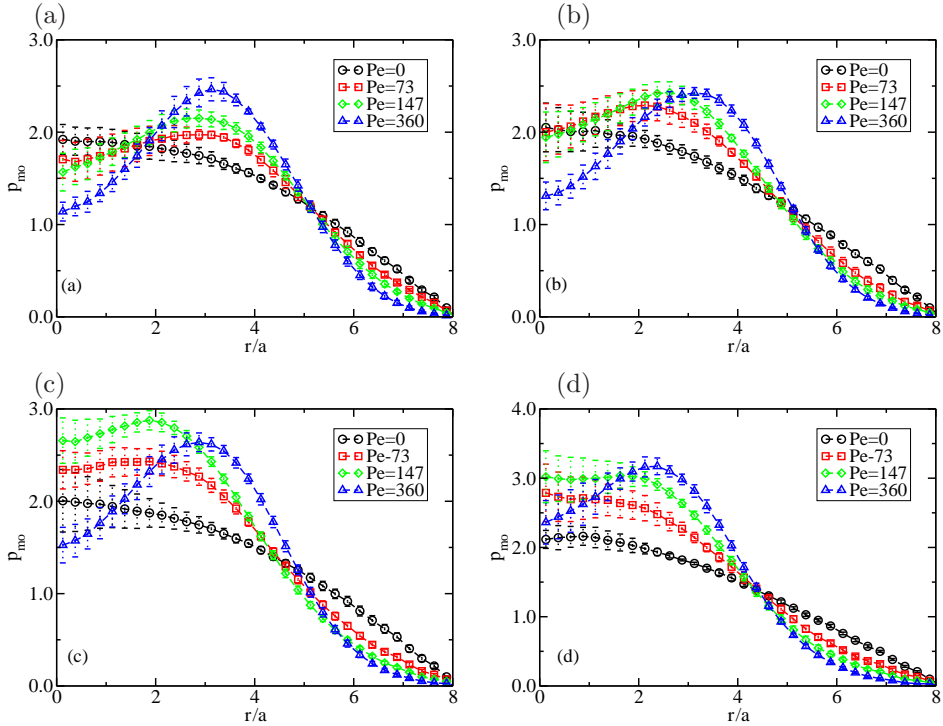

\begin{center}
\begin{tabular}{ll}
\quad(a)&\qquad(b)\\
\includegraphics*[width=60mm,clip]{fig2a.eps}&
\includegraphics*[width=60mm,clip]{fig2b.eps}\\
\quad(c)&\qquad(d)\\
\includegraphics*[width=60mm,clip]{fig2c.eps}&
\includegraphics*[width=60mm,clip]{fig2d.eps}
\end{tabular}
\caption{Radial monomer distributions for the Peclet numbers $Pe =0$
($\circ$), $73$ ($\square$), $147$ ($\diamond$), and $360$ ($\vartriangle$) of
polymers with $L_p/L_r \approx 3.8$  (a),  $7.7$ (b), $15.4$ (c),
and $30.8$ (d).
\label{fig:conc1}}
\end{center}
\end{figure}

The averages and probability distributions presented in the
following sections are calculated in the stationary state for
various independent initial conditions.

\section{Radial Distribution Functions}
\label{sec:concentration}

The imposed flow, with a parabolic flow profile, determines the
polymer conformations, with a few examples shown in
fig.~\ref{fig:snapshot}, and the polymer distribution functions,
such as the monomer distribution $P_{mo}$ and the center-of-mass
distribution $P_{cm}$. Their dependence on  flow strength will be
characterized in the following sections.

\subsection{Radial Monomer Distribution}

Radial monomer distributions are presented in fig.~\ref{fig:conc1}
for various flow rates and stiffnesses. They are normalized such
that
\begin{equation}
\int_0^{R/a} r P_{mo}(r/a) \ dr/a^2=1 .
\label{eqn2}
\end{equation}
A feature in common with all polymer systems with hydrodynamic
interactions is the decrease of the monomer concentration adjacent
to the wall with increasing flow rate. Even without flow, there is
a depletion zone for all stiffnesses, which extends approximately
one radius of gyration into the channel. However, the
distributions exhibit distinct differences to those obtained
previously for flexible and rodlike polymers, respectively. For
persistence lengths $L_p/L_r \lesssim 8$, the density at the
center of the channel decreases with increasing Peclet number as
shown in figs.~\ref{fig:conc1}a,b. In contrast, for stiffer
polymers with $L_p/L_r > 8$ the density at the channel center
first increases with increasing $Pe$ and then decreases at large
Peclet numbers. Simultaneously, the wall induced migration towards
the channel center causes an increase in concentration at a finite
distance from the center for all stiffnesses. Two effects
contribute to the formation of the maximum. On the one hand, there
is cross-streamline migration due to hydrodynamic interactions,
and, on the other hand, the flow field causes an alignment of a
molecule (see section~\ref{sec:alignment}), which increases the
local density at the radial positions of strong flow-induced
alignment.

\begin{figure}
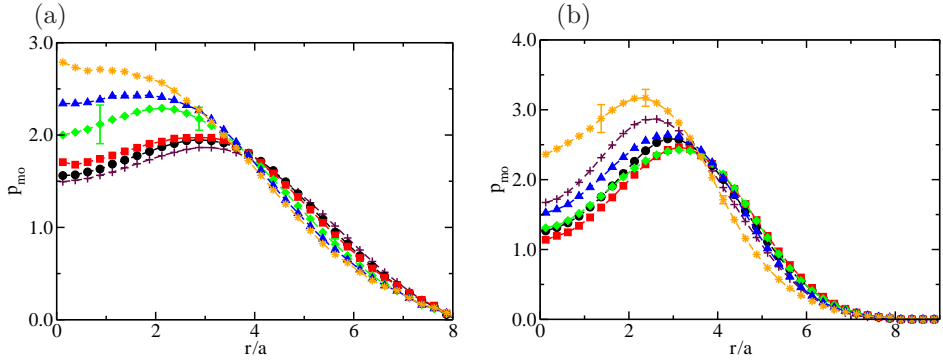

\begin{center}
\begin{tabular}{ll}
\quad(a)&\qquad(b)\\
\includegraphics*[width=60mm,clip]{fig3a.eps} &
\includegraphics*[width=60mm,clip]{fig3b.eps}
\end{tabular}
\caption{Radial monomer distributions for the Peclet numbers $Pe =73$ (a)
and $360$ (b) and the persistence lengths $L_p/L_r \approx 0.5$  ($+$),
$2.1$ ($\bullet$), $3.8$ ($\square$), $7.7$ ($\diamond$),
$15.4$ ($\vartriangle$), and $30.8$ ($\star$).
\label{fig:conc2} }
\end {center}
\end{figure}

The distribution functions are rather similar for the various
stiffnesses at $Pe=0 $ within the accuracy of the simulation.
Interestingly, stiffer polymers exhibit a more pronounced
migration away from the wall for $Pe \lesssim 100$ as is shown in
fig.~\ref{fig:conc2}a. Simultaneously, $P_{mo}$ decreases in the
channel center for small stiffnesses, leading to an off-center
maximum, and increases for larger ones, where the maximum is at
the center for some of the larger stiffnesses. With increasing
flow rate, off-center density maxima are obtained for all
stiffnesses. For the flow rate $Pe=360$, we observe a decrease in
the depletion zone adjacent to the wall with increasing stiffness
for $L_p/L_r < 8$ and a reversion of the trend at larger
stiffnesses as depicted in fig.~\ref{fig:conc2}b.

\begin{figure}
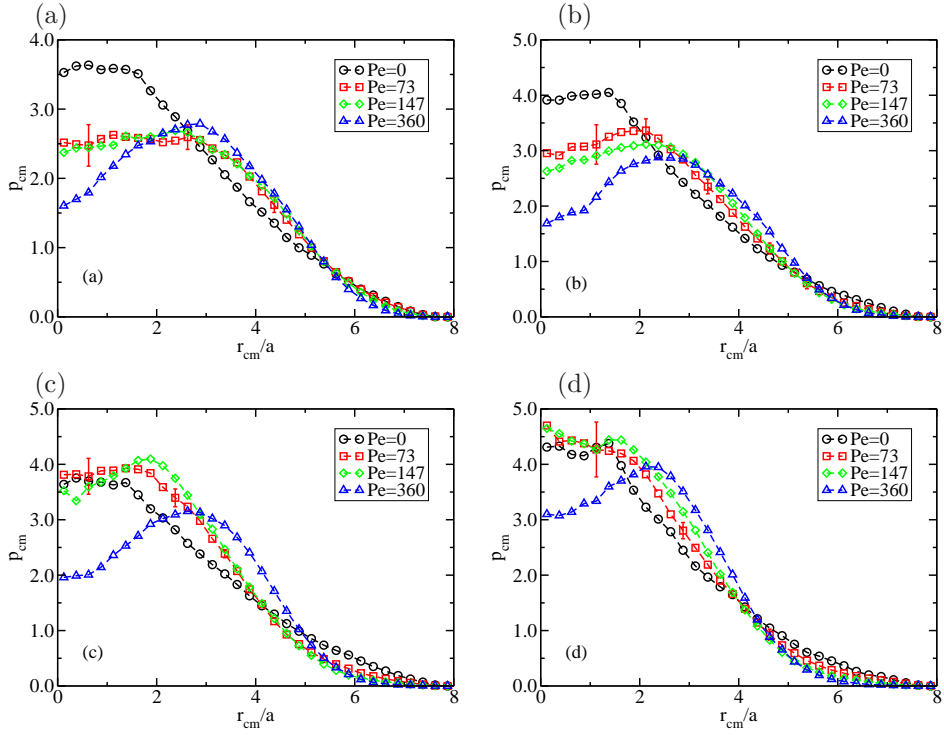

\begin{center}
\begin{tabular}{ll}
\quad(a)&\qquad(b)\\
\includegraphics*[width=60mm,clip]{fig4a.eps}&
\includegraphics*[width=60mm,clip]{fig4b.eps}\\
\quad(c)&\qquad(d)\\
\includegraphics*[width=60mm,clip]{fig4c.eps}&
\includegraphics*[width=60mm,clip]{fig4d.eps}
\end{tabular}
\caption{Radial center-of-mass distributions for the Peclet numbers $Pe =0$
($\circ$), $73$ ($\square$), $147$ ($\diamond$), and $360$ ($\vartriangle$) of
polymers with $L_p/L_r \approx 3.8$  (a),  $7.7$ (b), $15.4$ (c), and
$30.8$ (d).
\label{fig:cm1} }
\end{center}
\end{figure}

\begin{figure}
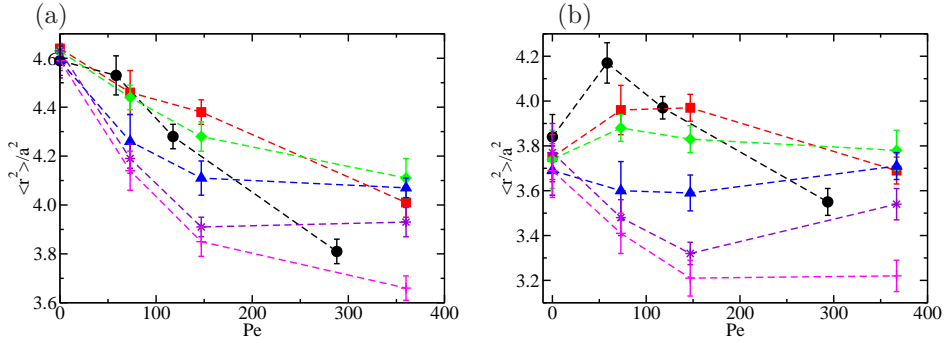

\begin{center}
\begin{tabular}{ll}
\quad(a)&\qquad(b)\\
\includegraphics*[width=60mm,clip]{fig6a.eps}&
\includegraphics*[width=60mm,clip]{fig6b.eps}
\end{tabular}
\caption{Widths $\langle r^2 \rangle$ of the radial monomer
distributions (a) and
center-of-mass distributions (b) for the persistence lengths
$L_p/L_r \approx 0.5$  ($\bullet$), $2.1$ ($\square$),
$3.8$ ($\diamond$), $7.7$ ($\vartriangle$), $15.4$ ($\star$),
and $30.8$ ($+$).
\label{fig:width} }
\end{center}
\end{figure}

The width of the distribution,
\begin{equation}
\langle r^2 \rangle = \int_0^{R/a} r^3 P_{mo}(r/a) \ dr/a^2 ,
\end{equation}
emphasizes the differences in migration behavior. As displayed in
fig.~\ref{fig:width}a, stiffer polymers exhibit a smaller width at
moderate Peclet numbers, indicating an enhanced net inward
migration with increasing bending rigidity, but $\langle r^2
\rangle$ seems to saturate at large flow rates. In contrast, the
widths of more flexible polymers show only minor changes at small
$Pe$ and decrease more rapidly at larger Peclet numbers. Their
widths can be smaller than those of stiffer polymers.

\begin{figure}
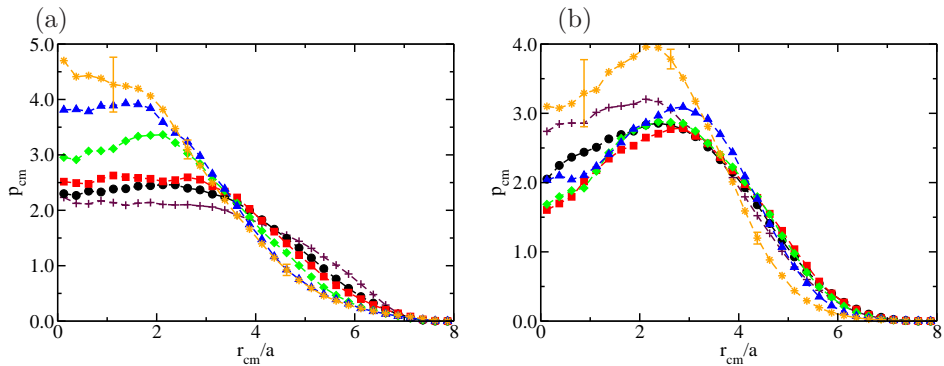

\begin{center}
\begin{tabular}{ll}
\quad(a)&\qquad(b)\\
\includegraphics*[width=60mm,clip]{fig5a.eps} &
\includegraphics*[width=60mm,clip]{fig5b.eps}
\end{tabular}
\caption{
Radial center-of-mass distributions for the Peclet numbers $Pe =73$ (a)
and $360$ (b) and persistence lengths $L_p/L_r \approx 0.5$  ($+$),
$2.1$ ($\bullet$), $3.8$ ($\square$), $7.7$ ($\diamond$),
$15.4$ ($\vartriangle$), and $30.8$ ($\star$).
\label{fig:cm2} }
\end {center}
\end{figure}

\subsection{Radial Center-of-Mass Distribution}

Radial polymer center-of-mass distributions are displayed in
fig.~\ref{fig:cm1}. Qualitatively, $P_{cm}$ exhibits similar
features as the monomer distribution $P_{mo}$, reflecting the same
physical mechanisms. For $L_p/L_r \lesssim 8$, the concentration
at the channel center decreases with increasing $Pe$. The density
profiles for $50 \lesssim Pe \lesssim 150$ are very similar and
indicate a weak dependence of the center-of-mass properties on the
flow rate only. For high $Pe$ values, the densities at the channel
center decrease significantly for all stiffnesses, and a clear
maximum appears at $r_{cm}> 0$. The density decrease with flow
near the wall is mainly due to wall induced cross-streamline
migration of polymers and partly due to steric interactions of the
very stiff polymers with the wall, while the density decrease at
the channel center indicates a migration away from the center. The
steric contribution to migration is evident from the Brownian
MPC simulations presented in sec.~\ref{sec:whi}. The two
competing migration mechanisms result in the formation of a
maximum at a distance away from the center at large $Pe$. The
maximum is less pronounced for the center-of-mass distribution
than for the monomer distribution.

Figure~\ref{fig:cm2} shows center-of-mass distributions of
polymers with different bending rigidities for $Pe \approx 73$ and
$Pe \approx 360$. As for even smaller Peclet numbers, the
distributions exhibit a maximum essentially in the channel center
for $Pe \approx 73$. An off-center maximum only appears for larger
flow rates, which is accompanied by a density decrease in the
channel center. Moreover, a pronounced depletion layer is visible
at the wall for large stiffnesses.

Flexible polymers exhibit a pronounced density increase adjacent
to the wall with increasing flow rate and before migration sets
in, whereas very stiff polymers exhibit an increased depletion
layer at the wall with increasing flow rate, which we attribute to
steric polymer-wall interactions. This is reflected in the flow
rate dependence of the width of the center-of-mass distribution
displayed in fig.~\ref{fig:width}b. For $L_p/L_r \lesssim 4$, the
width increases with increasing Peclet numbers for small $Pe$.
This increase is explained by the suppression of polymer-wall
steric interactions due to flow alignment of polymers (cf.
section~\ref{sec:alignment}). For flexible polymers such an
alignment is enhanced by flow-induced polymer stretching along the
channel axis. However, at larger $Pe$, the wall induced
hydrodynamic lift force dominates the dynamics, which results in a
decrease of width. This dependence is qualitatively different for
polymers with $L_p/L_r \gtrsim 5$, where wall induced migration
dominates even at low $Pe$, implying a decrease in width. However,
at large Peclet numbers, we observe a saturation or even increase
in width by outward migration due to hydrodynamic interactions,
which counter balances the radially inward migration, as
discussed in Sec. 6.

\section{Alignment and Conformations}
\label{sec:alignment}

\subsection{Orientational Order Parameter}

\begin{figure}
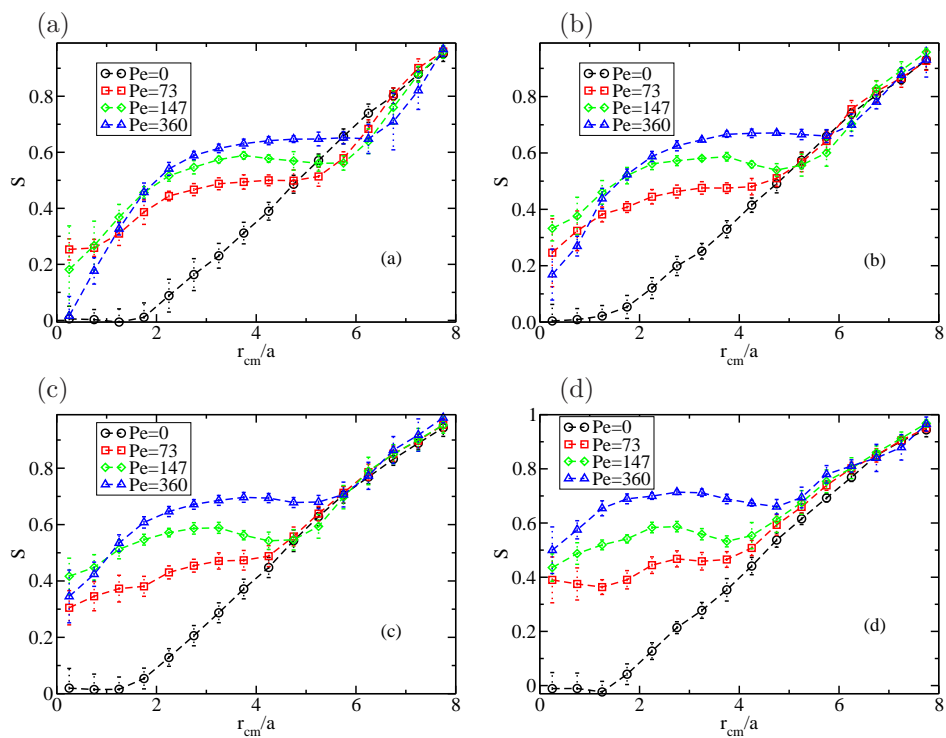

\begin{center}
\begin{tabular}{ll}
\quad(a)&\qquad(b)\\
\includegraphics*[width=60mm,clip]{fig7a.eps}&
\includegraphics*[width=60mm,clip]{fig7b.eps}\\
\quad(c)&\qquad(d)\\
\includegraphics*[width=60mm,clip]{fig7c.eps}&
\includegraphics*[width=60mm,clip]{fig7d.eps}
\end{tabular}
\caption{Orientation order parameter for the Peclet numbers $Pe =0$
($\circ$), $73$ ($\square$), $147$ ($\diamond$), and $360$ ($\vartriangle$) of
polymers with $L_p/L_r \approx 3.8$  (a),  $7.7$ (b), $15.4$ (c),
and $30.8$ (d).
\label{fig:op1} }
\end{center}
\end{figure}

\begin{figure}
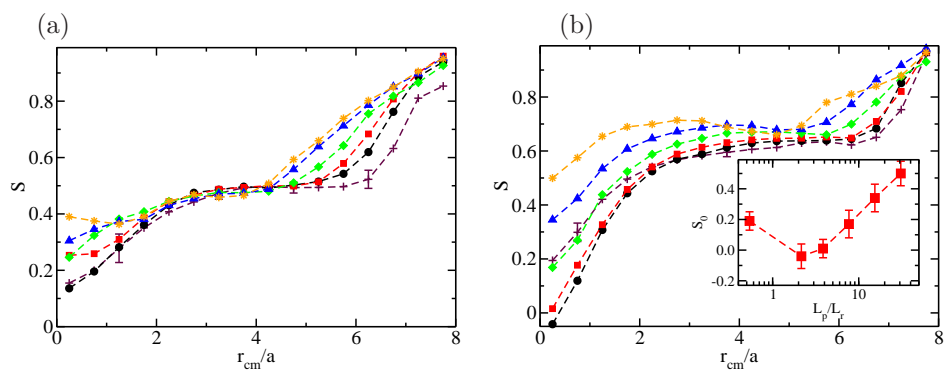

\begin{center}
\begin{tabular}{ll}
\quad(a)&\qquad(b)\\
\includegraphics*[width=60mm,clip]{fig8a.eps} &
\includegraphics*[width=60mm,clip]{fig8b.eps}
\end{tabular}
\caption{Orientational order parameters $S$ for Peclet numbers $Pe =73$ (a)
and $360$ (b) and persistence lengths $L_p/L_r \approx 0.5$  ($+$),
$2.1$ ($\bullet$), $3.8$ ($\square$), $7.7$ ($\diamond$),
$15.4$ ($\vartriangle$), and $30.8$ ($\star$).
Inset: $S$ close to the channel center for $Pe=360$.
The line is a guide to the eye.
\label{fig:op2}}
\end {center}
\end{figure}

In microchannel flows with no-slip boundary conditions the local
shear rate changes linearly with radial position. Hence, the
flow-induced force experienced by a polymer depends on its radial
position. Its conformations and alignment, in response to the flow
forces, is expected to vary with its bending rigidity and its
radial center-of-mass position. Studies of flexible and
semiflexible polymers~\cite{cann:08,chel:10} reveal large
orientational changes by the imposed flow, which is important for
cross-streamline migration~\cite{send:08}. To study the influence
of flow on the polymer orientation, we consider the orientational
order parameter
\begin{equation}
S(r_{cm})=\frac{1}{2}\langle 3 \cos^{2}\theta - 1 \rangle ,
\end{equation}
where $\theta$ is the angle between the polymer end-to-end vector
and the flow direction, as a function of its center-of-mass radial
position. $S(r_{cm})$ is plotted for various Peclet numbers
and stiffnesses in fig.~\ref{fig:op1}. In the absence of flow and
for all $\kappa_b$, the polymer orientation is isotropic at the
channel center. For distances close to the wall, confinement
causes a preferred orientation along the channel axis with $S
\approx 1$. With increasing flow rate, the order parameter
increases for all stiffness and reaches a plateau at large $Pe$
for certain radii. The plateau value and its extension depends on
stiffness as shown in fig.~\ref{fig:op2}.

The order parameter exhibits a non-monotonic dependence on  flow
strength close to the channel center and depends on polymer
stiffness as shown in the inset of fig.~\ref{fig:op2}. An increase
in stiffness leads to a decrease of the order parameter for
$L_p/L_r < 15$. For larger stiffnesses, $S(0)$ is larger
for larger $L_p/L_r$ at all radial distances. At $Pe=360$, $S$
decreases with increases stiffness and becomes even negative
for $L_p/L_r \simeq 2$, compare the inset of fig.~\ref{fig:op2}.
As the polymer stiffness increases, $S$ increases close to the channel
center. The negative values of $S$ at the channel center are
attributed to the formation of U-shaped conformations, compare
fig.~\ref{fig:snapshot}. Such
conformations are not possible for flexible polymers \cite{cann:08}
and polymers with large bending rigidity, hence $S$ assumes
higher values at the center as $\kappa_b$ increases.

\begin{figure}
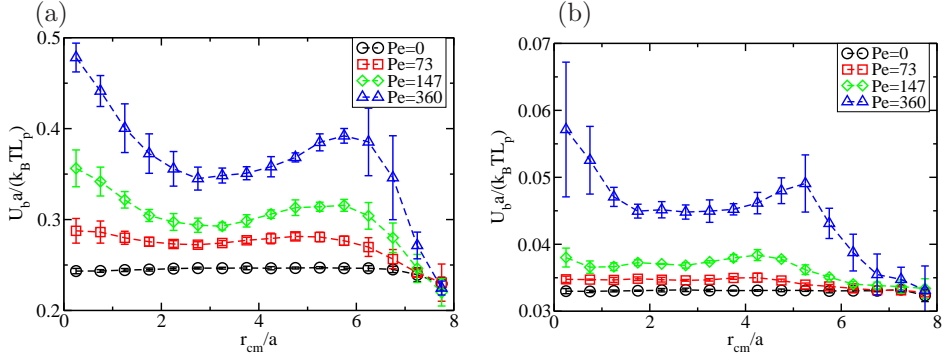

\begin{center}
\begin{tabular}{ll}
\quad(a)&\qquad(b)\\
\includegraphics*[width=60mm,clip]{fig9a.eps} &
\includegraphics*[width=60mm,clip]{fig9b.eps}
\end{tabular}
\caption{Bending energies $U_b$ for Peclet numbers $Pe =0$
($\circ$), $73$ ($\square$), $147$ ($\diamond$), and
$360$ ($\vartriangle$) of polymers with persistence lengths
$L_p/L_r \approx 3.8$ (a) and $30.8$ (b).
\label{fig:ub1}}
\end{center}
\end{figure}

\subsection{Bending Energy}

\begin{figure}
\begin{center}
\includegraphics*[width=60mm,clip]{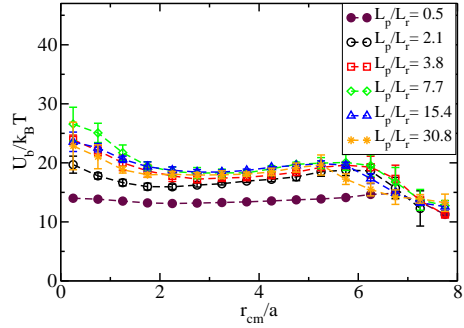}
\caption{Bending energies for Peclet number $Pe=360$ and persistence lengths
$L_p/L_r \approx 0.5$ ($\bullet$), $2.1$ ($\square$), $3.8$ ($\diamond$),
$7.7$ ($\vartriangle$), $15.4$ ($\star$), and $30.8$ ($+$).
\label{fig:ub2}}
\end {center}
\end{figure}

U-shaped conformations can be quantitatively analyzed by
calculating the average bending energy $U_{B}$.
Figure~\ref{fig:ub1} displays $U_b$ as  function of the radial
polymer center-of-mass position for various persistence lengths
and for various Peclet numbers. For $Pe = 0$ the bending energy is
nearly uniform across the channel cross section and decreases near
the wall, due to wall induced alignment. The magnitude of $U_b$ is
close to the thermal average $U_b = (L_p/a-1)k_B T$ of the nearly
harmonic bending potential for $L_p/L_r > 1$, as expected. An
increase in $Pe$ results in an increase in its absolute value,
which is a result of the conformational changes of the polymer in
response to the force exerted by the flow, compare
fig.~\ref{fig:snapshot}. At a given $Pe$ and
radial position, the flow exerts the same force on a polymer, but
its conformations depend on its bending rigidity. Polymers with
large bending rigidity hardly undergo conformational changes,
while polymers with rather small $\kappa_b$ go through significant
conformational changes, but their bending energies are similar.
This is exemplified in fig.~\ref{fig:ub2}, where $U_b/(k_BT)$ is
shown for various bending rigidities and $Pe = 360$.

In figs.~\ref{fig:ub1} and \ref{fig:ub2}, the radial energy
profiles are qualitatively similar for all $\kappa_b$ (with $L_p/L_r
>1$) for a given $Pe$. For $0.5 \lesssim L_p/L_r \lesssim 5$,
the polymer adopts U-shaped conformations at the center and hence
$U_b$ is large. Away from the center, $U_b$ decreases as U-shapes
disappear. With increasing radial center-of-mass position, the
local shear rate increases and a polymer with small persistence
length assumes transient bent conformations. The average over
individual configurations provides a high value for $U_b$ for such
$r_{cm}$. The extent of bending decreases as $\kappa_b$ is
increased, since for stiffer polymers $(1- \cos \vartheta) \sim
U_b/\kappa_b$, where $\vartheta$ is the angle between successive
bond vectors. More aligned bonds and hence smaller angle
$\vartheta$ imply that $U_b/\kappa_b$ decreases with increasing
$\kappa_b$, in agreement with the results of fig.~\ref{fig:ub1}.

\begin{figure}
\begin{center}
\begin{tabular}{ll}
\quad(a)&\qquad(b)\\
\includegraphics*[width=60mm,clip]{fig11a.eps}&
\includegraphics*[width=60mm,clip]{fig11b.eps}
\end{tabular}
\caption{Radial monomer distributions for systems without hydrodynamic
interactions.
The persistence lengths are $L_p/L_r = 3.8$ (a) and $30.8$ (b), and the
Peclet numbers $Pe =0$ ($\circ$), $147$ ($\square$),
$360$ ($\vartriangle$), and $920$ ($\star$).
\label{fig:dis_mon_nohi}}
\end{center}
\end{figure}

\begin{figure}
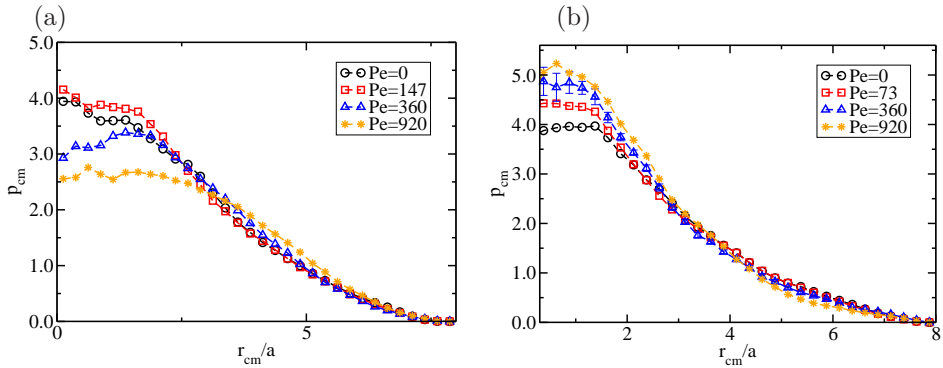

\begin{center}
\begin{tabular}{ll}
\quad(a)&\qquad(b)\\
\includegraphics*[width=60mm,clip]{fig12a.eps}&
\includegraphics*[width=60mm,clip]{fig12b.eps}
\end{tabular}
\caption{Radial center-of mass distributions for systems without
hydrodynamic interactions.
The persistence lengths are $L_p/L_r = 3.8$ (a) and $30.8$ (b) and the
Peclet numbers $Pe =0$ ($\circ$), $147$ ($\square$), $360$ ($\vartriangle$),
and $920$ ($\star$).
\label{fig:dis_cm_nohi} }
\end{center}
\end{figure}

\section{Structural Properties without Hydrodynamic Interactions}
\label{sec:whi}

As is generally accepted by now, hydrodynamic interactions
determine the behavior of polymers in microcapillary flows. In
order to separate effects due to intramolecular and surface
hydrodynamic interactions from those caused by the flow and steric
interactions, we performed Brownian MPC simulations
\cite{ripo:07}. Brownian dynamics simulation have also been
performed in Ref.~\cite{sain:06}, with a polymer composed of
rodlike segments. However, intramolecular hydrodynamic
interactions are taken into account by employing anisotropic drag
coefficients parallel and perpendicular to the rodlike segments.
In contrast, our polymer model is composed of point-like monomers
with isotropic friction. Hence, we consider a free draining chain,
whereas in Ref.~\cite{sain:06} intramolecular hydrodynamic
interactions are taken into account implicitly.

\subsection{Radial Concentration Distributions}

The radial monomer probability distributions for the persistence
lengths $L_p/L_r \approx 2.1$ and $30.8$ are presented in
fig.~\ref{fig:dis_mon_nohi} for various flow rates. For $Pe <
200$, the distribution functions are rather similar. Adjacent to
the wall, the distributions for small stiffnesses are even similar
for all flow rates. At large flow rates and persistence lengths
$L_p/L_r < 3.5$, we find a monomer density increase at $r_{cm}/a \approx 4$,
which we attribute to the induced alignment of the polymers by the
flow. However, for larger stiffnesses, we find an inward migration
adjacent to the wall due to increased steric polymer-surface
interactions (cf. fig.~\ref{fig:dis_mon_nohi}b). Here, the polymer
tumbling motion combined with the smaller number of conformational
degrees of freedom at large stiffnesses leads to a strong
repulsion from the wall. At the same time, the density in the
channel center increases with a maximum at $r\approx 0$.

The radial center-of-mass distribution exhibits an outward
migration even without hydrodynamic interactions for $L_p/L_r <
3.5$ due to polymer alignment (cf. fig.~\ref{fig:dis_cm_nohi}). In
contrast, $P_{cm}$ becomes narrower with increasing flow rate for
$L_p/L_r \approx 30.8$, similar to the monomer distribution.
Hence, a qualitative different behavior is obtained depending on
the ability of the polymer to adjust to the conformational
restrictions by the walls.

The Brownian simulations of ref.~\cite{sain:06} display in
strong outward migration of the polymers, which leads to a
density minimum at the channel center and a large density increase
near the walls. Two effects contribute to this behavior. On the
one hand alignment of the polymers and on the other hand
anisotropic diffusion by the aligned segments. We do not observe
such an increase in density adjacent to the wall, because no
hydrodynamic effects are included in our simulations.

\subsection{Alignment}

The orientational order parameters display qualitatively similar
dependencies on flow rate as those for systems with hydrodynamic
interactions (cf. figs.~\ref{fig:op1} and \ref{fig:op2}). However,
alignment is far less pronounced and much larger Peclet numbers
are required to achieve  significant flow alignment. For $Pe=360$,
we find $S\approx 0.2$ at $L_p/L_r \approx 3.8$, compared to
$S=0.6$ in the presence of hydrodynamic interactions. At the same
time, we also observe the appearance of U-shaped structures in the
channel center for larger Peclet numbers. The comparison shows
that certain qualitative features of polymer alignment are
determined by the flow profile, rather than hydrodynamic
interactions. However, hydrodynamic interactions clearly influence
the polymer orientation in a quantitative manner and lead to
a more pronounced alignment.

\section{Cross-Streamline Dynamics}
\label{sec:migration}

The most striking effect of flow on a semiflexible polymer is
the strong dependence of its orientation on $Pe$ and the radial
distance. This aspect provides the key to understand the
appearance of the off-center maximum in the center-of-mass
distribution function.

As is well known, a rod in solution exhibits a larger diffusion
coefficient parallel to its axis than perpendicular to it. In the
absence of flow, a semiflexible polymer behaves very similar to
a stiff rod, we therefore expect that the observed polymer
orientational differences across the channel will lead to
differences in the lateral diffusion behavior.

\begin{figure}
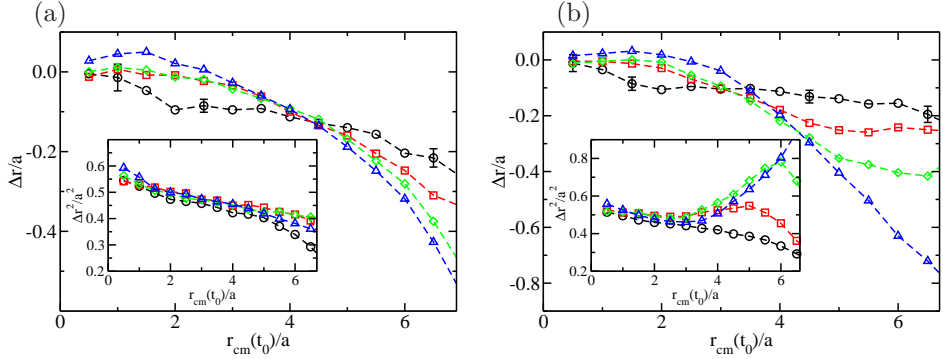

\begin{center}
\begin{tabular}{ll}
\quad(a)&\qquad(b)\\
\includegraphics*[width=60mm,clip]{fig13a.eps} &
\includegraphics*[width=60mm,clip]{fig13b.eps}
\end{tabular}
\caption{Radial center-of-mass displacements $\Delta r (r_{cm})$ and
mean square displacements $\Delta r^2(r_{cm})$ (insets) for Peclet
numbers $Pe =0$
($\circ$), $73$ ($\square$), $147$ ($\diamond$), and $360$ ($\vartriangle$) of
polymers with persistence lengths $L_p/L_r \approx 3.8$ (a) and $30.8$ (b).
\label{fig:drift_stiffness}}
\end{center}
\end{figure}

\begin{figure}
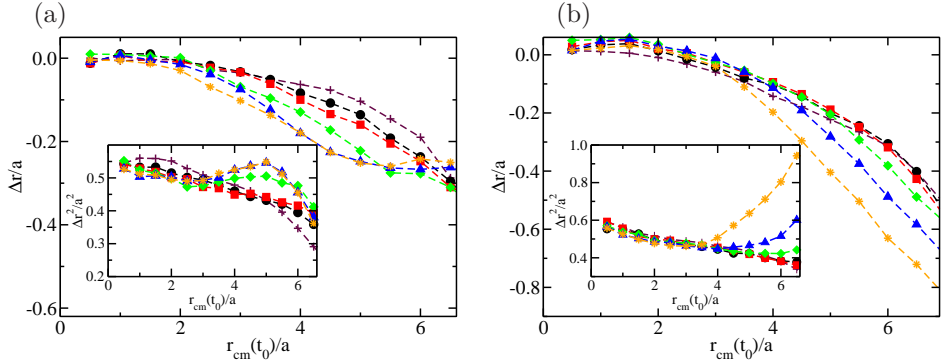

\begin{center}
\begin{tabular}{ll}
\quad(a)&\qquad(b)\\
\includegraphics*[width=60mm,clip]{fig14a.eps} &
\includegraphics*[width=60mm,clip]{fig14b.eps}
\end{tabular}
\caption{Radial center-of-mass displacements $\Delta r (r_{cm})$ and
mean square displacements $\Delta r^2(r_{cm})$ (insets) for
persistence lengths $L_p/L_r \approx 0.5$  ($+$), $2.1$ ($\bullet$),
$3.8$ ($\square$), $7.7$ ($\diamond$), $15.4$ ($\vartriangle$),
and $30.8$ ($\star$), and $Pe=73$ (a) and $Pe=360$ (b).
\label{fig:drift_peclet}}
\end{center}
\end{figure}

In order to characterize this diffusive behavior, we calculate the
radial mean displacement $\Delta r (r_{cm}) = \left\langle [{\bm
r}_{cm}(t_0 + \Delta t) - {\bm r}_{cm}(t_0)] {\bm e}_{cm} (t_0)
\right\rangle$, where ${\bm e}_{cm}= {\bm r}_{cm}/ |{\bm
r}_{cm}|$, and the center-of-mass mean square displacement $\Delta
r^2(r_{cm}) = \left\langle ({\bm r}_{cm}(t_0 + \Delta t) {\bm
e}_{cm} (t_0) - \left\langle {\bm r}_{cm}(t_0) {\bm e}_{cm}
(t_0)\right\rangle )^2 \right\rangle$ within a certain time
$\Delta t$ for a polymer with the center-of-mass position  ${\bm
r}_{cm}$ at $t_0$. We choose the time $\Delta t$ such that the
center-of-mass displacement is smaller than $a$ and the mean
square displacement is linear in time.  For an homogeneous and
isotropic system, $\Delta r$ and $\Delta r^2$ are independent of
the initial position.

The mean values are the first and second moment of the conditional
probability distribution function to find the polymer
center-of-mass at position $r_{cm}(t_0+\Delta t)$ at the time $t_0
+ \Delta t$, when it has been at position $r_{cm}(t_0)$ at the
time $t_0$. Initially, the distribution is very narrow, ideally a
$\delta$ function, and broadens in time. In the limit $\Delta t
\to \infty$ the conditional probability distribution turns into
the equilibrium distribution function as displayed in
figs.~\ref{fig:cm1} and \ref{fig:cm2}.

Figures~\ref{fig:drift_stiffness} and \ref{fig:drift_peclet}
display $\Delta r$ and $\Delta r^2$ for various Peclet numbers and
stiffnesses. The mean displacements evidently depend on flow rate
and stiffness. At $Pe=0$, there is no drift in the channel center.
Only for $r_{cm}/a \gtrsim 1.5$, we observe a drift, which we
attribute to steric polymer wall interactions, and which increase
adjacent to the wall. For larger flow rates, the drift for $r_{cm}
\lesssim 4$ is close to zero due to alignment of a polymer and
lack of steric wall interactions (cf. fig.~\ref{fig:drift_peclet}
a). At large Peclet numbers, $\Delta r$ is positive for $r_{cm}/a
< 2.5$ (cf. fig.~\ref{fig:drift_peclet}b) and all persistence
lengths $L_p/L_r>2$, which indicates an outward migration. The
wall interactions lead to a larger drift of polymers towards the
channel center adjacent to the wall, which increases with
increasing stiffness.

The mean square displacement decreases with increasing radial
distance for $r_{cm} < 4 a$ at all stiffnesses and flow rates.
Since $\Delta r$ is rather small for $r_{cm} < 4 a$, $\Delta r^2$
is determined by diffusion, i.e., the radial diffusion is larger
in the channel center. This leads to a minimum in the
center-of-mass distribution at large flow rates and stiffnesses.
At larger radial distances and large stiffnesses, the mean square
displacement is strongly affected by wall interactions. There is
no simple explanation for the non-monotonic behavior of the mean
square displacement. Various wall interactions contribute to the
system behavior: the wall lift force, steric wall interactions,
and an increase of the tumbling frequency with increasing flow
rate \cite{chel:10,wink:06_1}.

\begin{figure}
\begin{center}
\includegraphics*[width=60mm,clip]{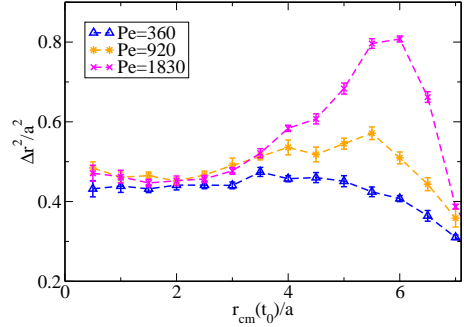}
\caption{Radial center-of-mass
mean square displacements without hydrodynamic interactions for the
persistence lengths $L_p/L_r \approx 30.8$ and the Peclet numbers
$Pe=360$ ($\vartriangle$), $920$ ($\star$), and $1830$ ($\times$).
\label{msd_nohi}}
\end{center}
\end{figure}

{\em Without hydrodynamic interactions}, the drift at $Pe=0$ is
comparable to that of the system with hydrodynamic interactions.
It changes  with increasing Peclet number due to flow alignment,
but only slightly. The mean square displacement is independent of
the radial position for distances $r_{cm} < 6 a$ for all Peclet
numbers $Pe <1000$ and stiffnesses as shown in
fig.~\ref{msd_nohi}. Only close to the wall, there is a decrease
due to steric wall interactions. This supports our conclusion that
intramolecular hydrodynamic interactions are responsible for
outward migration in systems with hydrodynamic interactions. At
much larger Peclet numbers ($Pe \approx 2000$), the mean square
displacement is constant for $r<4a$, increases then with the
radial distance and decreases again beyond $r =6 a$, similar to
the behavior shown in the inset of fig.~\ref{fig:drift_peclet}a
for $L_p/L_r =15.4$. This behavior is explained by the strong
steric polymer wall interaction and fast tumbling motion at high
flow rates. Note that the tumbling frequency increases with the
shear rate \cite{chel:10,wink:06_1}. Hence, systems with and
without hydrodynamic interactions exhibit similar features, but
larger Peclet numbers are required for systems without
hydrodynamic interactions and the behavior of systems with
hydrodynamic interactions is richer and additional features are
displayed.

\section{Summary and Conclusions}

We have analyzed the flow behavior of semiflexible polymers
confined in a cylindrical channel and systematically studied the
influence of bending rigidity on structural properties and
cross-streamline migration. The study includes polymers with
moderate bending rigidity, comparable to actin filaments, to very
stiff rodlike polymers.

Poiseuille flow induces a  strong radial-dependent
polymer alignment parallel to the channel associated with  a
radially outward migration of its center-of-masses at small Peclet
numbers and, at large Peclet numbers, a wall induces inward
migration.

Stiffness changes the polymer behavior qualitatively and
quantitatively. Taking a system {\em without} hydrodynamic
interactions as a reference, there is no outward migration and
there are no wall hydrodynamic forces. Correspondingly, the
center-of-mass density decreases in the channel center with
increasing flow rate and up to a certain stiffness ($L_p/L_r <20$)
due to polymer alignment. For large stiffnesses, however, steric
polymer-wall interactions cause inward migration of the molecule
and an increase in density at the channel center with increasing
flow rate. We attribute this effect to the increase of the
tumbling frequency at large flow rates and hence more polymer-wall
contacts. Hence, even for such a system, a qualitative different
behavior is found for large and small stiffnesses.

This picture is enriched by hydrodynamic interactions, which
enhance outward migration. At the same time wall-lift forces lead
to inward migration and an off-center density maximum is formed at
large flow rates. Such a maximum has also been found in
experiments using actin filaments \cite{stei:08}. The
center-of-mass distribution strongly depends on stiffness and the
presence of hydrodynamic interactions, as is reflected in its
width. Flexible polymers exhibit a more complex migration behavior
than stiffer polymers due to conformational changes, i.e,
stretching and alignment. For large stiffnesses, wall steric
interactions lead to a strong inward migration as for polymers
without hydrodynamic interactions. The semiflexible polymers
exhibit a non-universal behavior when changing stiffness and flow
rate. At low flow rates, stiffer polymers display a stronger
inward migration, whereas at high flow rates a reversed trend is
observed initially and ultimately a strong inward migration at
large stiffnesses. It is the interplay between confinement,
hydrodynamic interactions, and stiffness which leads to the
emergence of a complex structural and dynamical behavior of the
polymer. This is reflected in the stiffness dependence of
the mean radial and mean square radial displacements.

Although the understanding of the flow behavior of soft mesoscale
particles in nano- and microchannels has made considerable
progress in recent years, this field can be expected to attract
far more attention in the future. For example, flow in
structured channels has hardly been addressed so far. Also,
interactions between polymers at finite concentration, and between
polymers and other soft particles, needs to be investigated.

\subsection{Acknowledgments}
The financial support and the stimulating environment of the
DFG priority program ``Nano- and Microfluidics" is gratefully
acknowledged. In particular, we thank R.~Finken and U.~Seifert
(Stuttgart), L.~Schmid and T.~Franke (Augsburg), and D. Steinhauser and
T. Pfohl (G\"ottingen \& Basel) for many helpful and inspiring discussions.

\section*{References}
\bibliographystyle{unsrt}
\bibliography{polymer_microfluidics}

\end{document}